\documentclass{article}
\usepackage{authblk}
\usepackage[utf8]{inputenc}
\usepackage{graphicx}
\usepackage{bm}
\usepackage{dcolumn}
\usepackage{amssymb}
\usepackage{amsmath}

\title{Density wave ground state and fractional fermions in $LaAlO_3/SrTiO_3$\ \  heterostructure}

\author[1]{Sohail Dasgupta \thanks{sohaildasgupta@gmail.com}}
\author[2] {Vivek M. Vyas \thanks{physics.vivek@gmail.com}} 
\author[1]{Prasanta K. Panigrahi \thanks{pprasanta@iiserkol.ac.in}}

\affil[1] {Indian Institute of Science Education and Research Kolkata, Mohanpur - 741246}
\affil[2] {Raman Research Institute, Bangalore- 560080}

\begin{document}
\maketitle

\textbf{Abstract - }We show that the non-homogeneous charged layer distribution of the $LaAlO_3/SrTiO_3$ heterostructure undergoing interface reconstruction is the density wave ground state of the well known anharmonic lattice model described by the $\lambda \phi^4$ continuum model.
The two dimensional planar structure of the charged surfaces with alternating polarity leads to an effective one dimensional model, with fermions coupled to the planar distortions acting as long wavelength optical phonons in one dimension.
The Hamiltonian with the desired anharmonicity for describing the non-homogeneous density wave type distortion is the same one that describes the fermion number fractionalization in polyacetylene.
The general solution of this theory is the Jacobi elliptic sine function sn(x;k), which in the limiting case of lattice distortion being localized gives the kink/anti-kink solution.

\hrulefill
\\
Charge fractionalization has been observed in a heterostructure comprising of $LaAlO_3$ and $SrTiO_3$ \cite{ohtomo2004high,nakagawa2006some}, having alternate $(LaO)^+$ and $(AlO_2)^-$ atomic planes with $SrTiO_3$ acting as the non-polar substrate.
The non-zero electric field between the oppositely charged atomic planes of $LaAlO_3$ creates an in-built electrostatic potential that diverges with the thickness.
This polar discontinuity at the interface forces a charge redistribution with the end layers obtaining a net charge of magnitude $\frac{e}{2}$.
The 2D electron gas formed at the interface of these oxide heterostructures have been observed to have many interesting properties such as ferromagnetism \cite{lee2013titanium,brinkman2007magnetic}, metallic conductivity \cite{ohtomo2004high, thiel2006tunable, caviglia2008electric, bell2009dominant}, superconductivity \cite{ueno2008electric}, coexistence of magnetic order and superconductivity \cite{bert2011direct}, electron phase separation at the interface \cite{wang2011electronic} and room temperature photo-conductivity \cite{tebano2012room}, thus creating possibility of many physical applications such as sensors, photo-detectors \cite{bogorin2010laalo3} and solar cells \cite{assmann2013oxide}.
Very recently the elusive two dimensional hole gas has also been observed in this interface \cite{lee2018direct,chen20182d}, opening up further prospects of real life applications.

Generally, the oxide heterostructure is made up of two different $ABO_3$ type perovskites, having alternating $AO$ and $BO_2$ atomic planes along the $(001)$ orientation. 
$LaAlO_3$ is a $A^{3+}B^{3+}O_3$ perovskite with alternating $(LaO)^+$ and $(AlO_2)^-$ atomic planes whereas $SrTiO_3$ is a $A^{2+}B^{4+}O_3$ perovskite with alternating $(SrO)^0$ and $(TiO_2)^0$ atomic planes acting as a non-polar substrate over which the polar $LaAlO_3$ is stacked layer by layer.
There exists a non-zero electric field between the charged layers with the corresponding potential diverging as a function of layer thickness.
A net electron (hole) transfer per unit cell to the nearest neighbor is energetically favored, causing the electrostatic potential to oscillate between the layers, stabilizing the system with the end layers obtaining a net $\frac{e}{2}$ charge of opposite polarity \cite{nakagawa2006some}.

This is the phenomenological polar catastrophe model which has been recently given a field theoretic description \cite{selvan2016charge}.
This work makes use of the Bell-Rajaraman (BR) model \cite{bell1983states} which explains fermion number fractionalization in the soliton free sector.
The BR model does not incorporate dynamical phonons present in the system, albeit incorporating phonon induced non uniformity in the density wave ground state, giving charge fracionalization as an end result.

Here we show that the common origin of the dynamics of the phonon field in both the systems is the self interacting $\lambda \phi^4$ theory.
The non-homogeneous ground state of the system is the general Jacobi elliptic sine function of the $\lambda \phi^4$ theory.
In the limiting case when the modulus parameter, $k$ equals unity, this yields the kink/anti-kink solution describing the localized lattice distortion appropriate for the polyacetylene case, which carries a topological charge of $Q=+/- 1$. 
Due to the conservation of topological charge , kink and anti-kink are produced in pairs, keeping the net topological charge to be zero.

\begin{figure}
\centering
\includegraphics[scale=0.07]{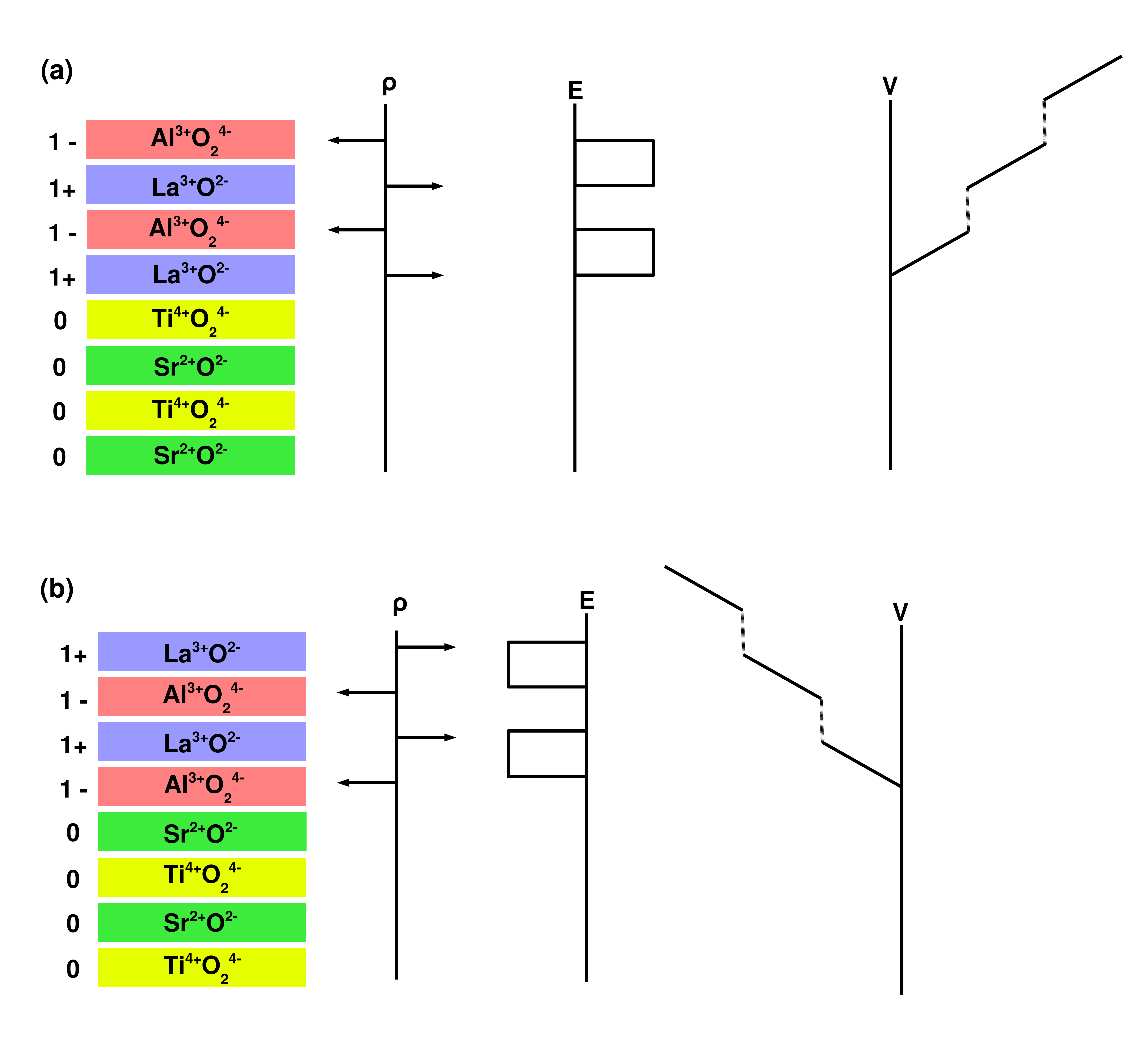}
\caption{(001) oriented LaAlO$_{3}$/SrTiO$_{3}$ heterostructure. (a) Unreconstructed (LaO)$^{+}$/(TiO$_{2}$)$^{0}$ interface.  (b) Unreconstructed (AlO$_{2}$)$^{-}$/(SrO)$^{0}$ interface.}
\label{unreconstructed}
\end{figure}

\begin{figure}
\centering
\includegraphics[scale=0.07]{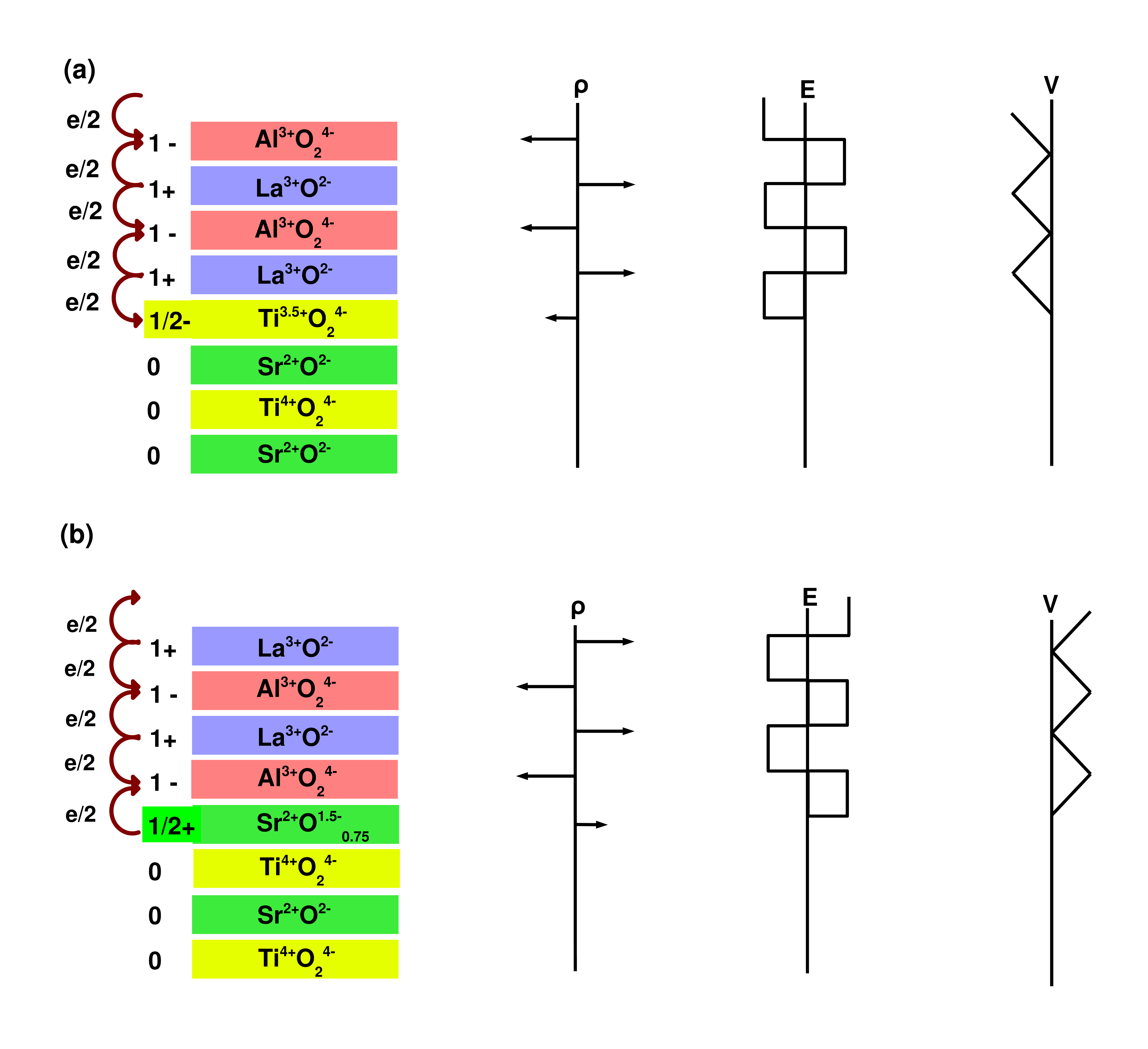}
\caption{Electron redistribution along the (001) orientation forced by polar discontinuity. (a) Half an electron per two dimensional unit cell is transferred in the case of (LaO)$^{+}$/(TiO$_{2}$)$^{0}$ interface.  (b) In (AlO$_{2}$)$^{-}$/(SrO)$^{0}$ interface, half a hole per two dimensional unit cell is transferred. }
\label{reconstructed}
\end{figure}

The $LaAlO_3/SrTiO_3$ heterostructure system, depicted in Fig.(\ref{unreconstructed}) has alternate layers of oppositely charged planes built on a neutral $SrTiO_3$ interface.
The corresponding dipolar electric field is a step function.
As the number of layers increases substantially, the potential diverges in the case of unreconstructed interface.
To prevent this polar catastrophe, an electron/hole gets shared by two unit cells causing the end layers to obtain a charge of $\frac{e}{2}$.
This makes the electric field to oscillate with an amplitude being half the constant value for the unreconstructed interface, making the potential also oscillatory after reconstruction as shown in Fig.(\ref{reconstructed}).

\begin{figure}
\centering
\includegraphics[scale=0.2]{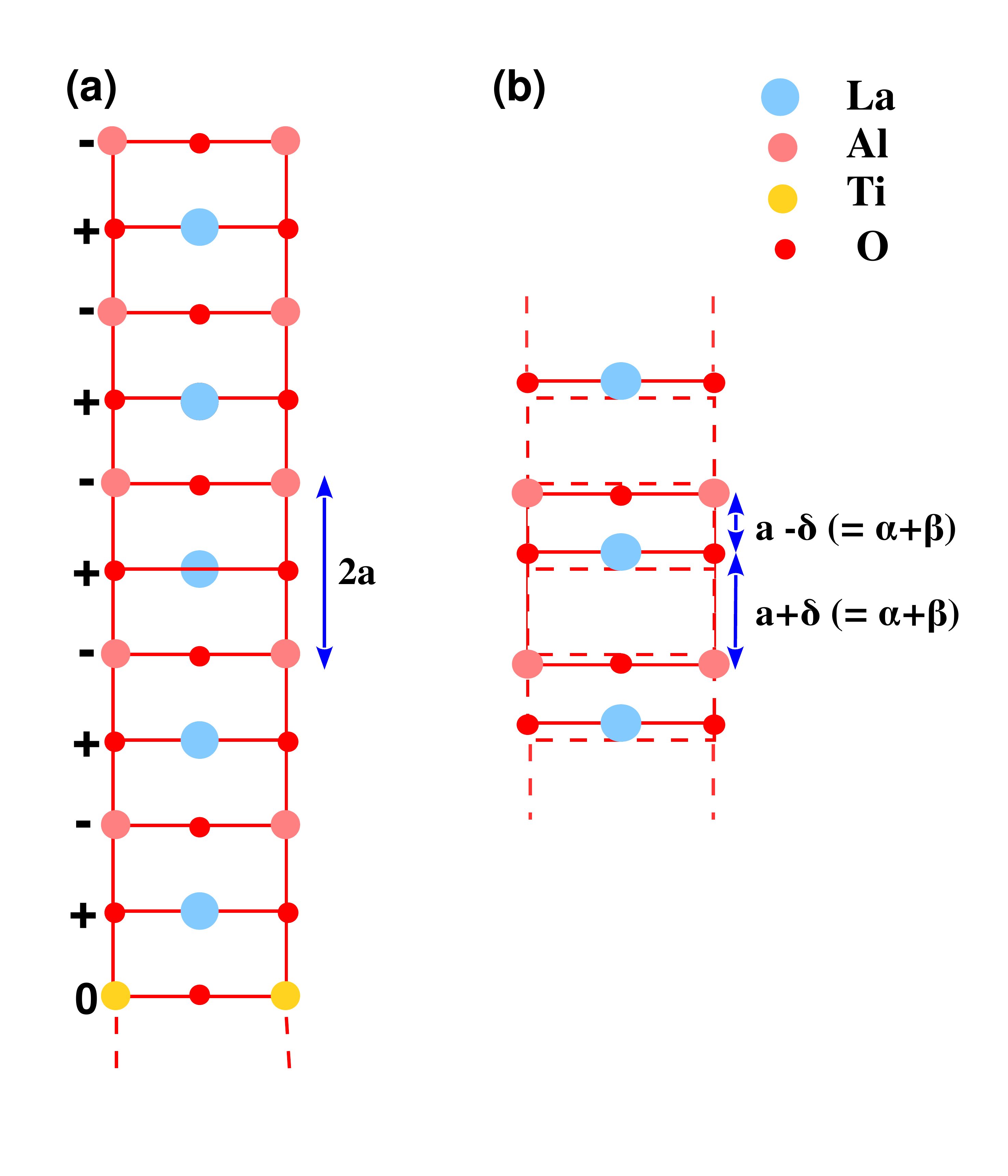}
\caption{(a) Two dimensional view of (LaO)$^{+}$/(TiO$_{2}$)$^{0}$ interface structure consisting of $2M$ alternatively charged (LaO)$^{+}$ and (AlO$_{2}$)$^{-}$ planes and a neutral (TiO$_{2}$)$^{0}$ plane. (b) Atomic planes are displaced from their initial position. (LaO)$^{+}$ plane is displaced by an amount ${\alpha}$ and (AlO$_{2}$)$^{0}$ planes are displaced by an amount ${\beta}$ (not specified here). The inter-planer distance changes by $(-)^{n}{\delta} (={\alpha} + {\beta})$ for adjacent planes.}
\label{2dview}
\end{figure}

\noindent
Since, the electron hopping is along the (001) direction, we can consider this system as quasi one dimensional.
As is well known, due to Peierls instability \cite{peierls1955quantum}, an equally spaced one dimensional chain becomes unstable for a non-zero electron-phonon coupling strength leading to chain distortions, thus creating a density wave in the system, wherein the electron density and the nuclear displacements oscillate in space.
The ground state is non-uniform; layers having a displacement from their equilibrium position with two neighboring layers coming close to each other, while the next two neighbors split apart by the same amount as shown in Fig.(\ref{2dview}).
This is caused by lattice anharmonicity \cite{kittel1953introduction,mukherjee1996thermal}.
As the system has an inherent charge conjugation symmetry, the anharmonicity cannot be of odd order and the simplest anharmonic term that can be added is of quartic type, enabling us to model it with the well known double well potential with the unreconstructed interface state corresponding to the $Z_2$ unbroken phase.
The anharmonicity inherently present in the system spontaneously breaks the $Z_2$ symmetry giving the general solution to the phonon field as the periodic Jacobi elliptic sine function, which describes the oscillatory density wave state of the reconstructed interface as depicted in Fig.(\ref{reconstructed}).

\noindent
This system is equivalent to a one dimensional polyacetylene chain with each  atomic plane representing a $CH$ group and the inter planar gap representing a chemical bond; the longer gaps equivalent to single and the shorter gaps to double bonds.
Polyacetylene has doubly degenerate ground states, exactly like the heterostructure system, with solitons to interpolate between them \cite{rao2008fermion}.
\begin{figure*}
\centering
\includegraphics[scale=.7]{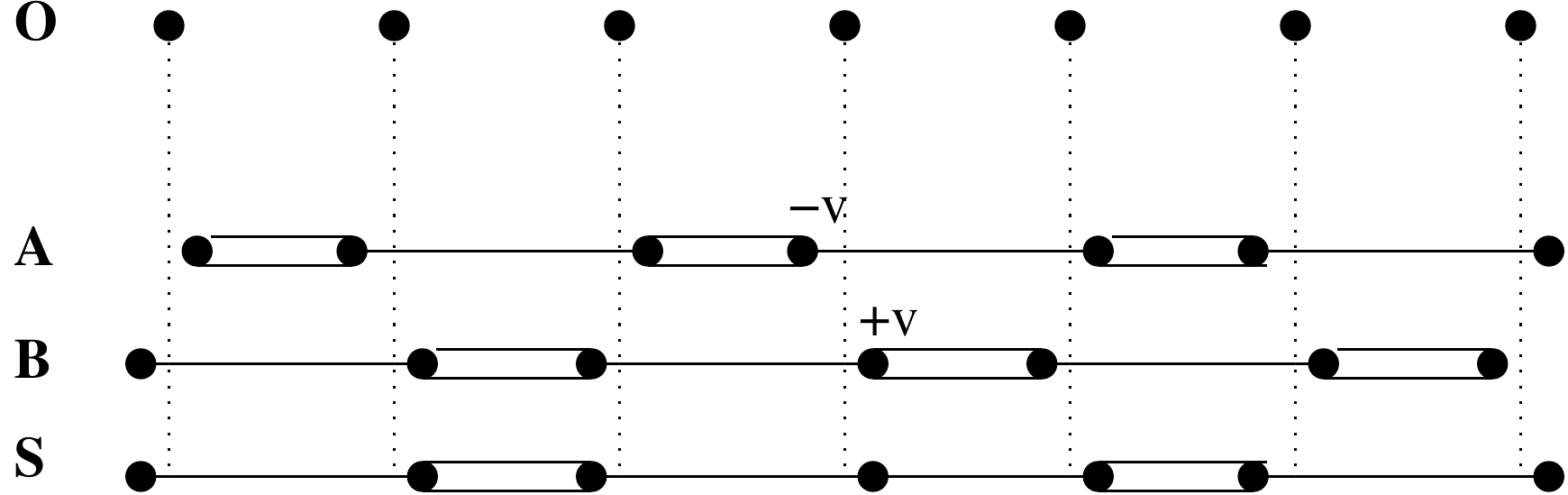}
\caption{Schematic of 1D polyacetylene chain.O is the lattice with equal lattice spacing. A and B are the two ground states formed due to Peierls instability in the equally spaced system. S is the soliton interpolation of the two ground states.}
\label{pol}
\end{figure*}

Jackiw and Rebbi first showed the existence of fractionally charged solitons \cite{jackiw1976solitons}.
Jackiw and Schreiffer demonstrated charge fractionalization \cite{jackiw1981solitons} in the soliton sector of polyacetylene using the Su-Schreiffer-Heeger (SSH) model \cite{su1979solitons}. 
Bell and Rajaraman exhibited fractionaly charged solitons \cite{rajaraman1982solitons} and fermion number fractionalization in the soliton-free sector using a simpler version of the SSH model.
Selvan and Panigrahi \cite{selvan2016charge} have used the BR model,

\begin{equation}\label{BR-model}
H = \sum \limits_n \Bigg(c^+(n+1)c(n)+c^+(n)c(n+1)\Bigg)\Bigg(u(n+1)-u(n)-\frac{1}{2a}\Bigg)
\end{equation}

to describe the charge fractionalization in $LaAlO_3/SrTiO_3$ heterostructure interface, giving a field theoretic description of the polar catastrophe model.
Here, $a$ is the lattice constant of the unreconstructed lattice, $u(n)$ represents a phonon field, and the $c(c^+)$ are fermion destruction (creation) operators at the lattice site $n$.
This Hamiltonian is a simplification of the model used by Jackiw and Schreiffer \cite{jackiw1981solitons}, which includes a spin variable and the dynamics of the coupled phonon field $u$.
Although the BR model suffices to describe the charge fracitonalization, it fails to take into account the origin of the lattice displacement due to the coupling of the phonon field.
We shall invoke the SSH Hamiltonian \cite{su1979solitons}:

\begin{equation}\label{ham}
H = \sum\limits_{n=1}^N\Big(\frac{p_n^2}{2m}+V(u_n,u_{n+1})\Big) - \sum\limits_{n=1,s=\pm\frac{1}{2}}^N t_{n+1,n}(c_{n+1,s}^\dagger c_{n,s}+c_{n,s}^\dagger c_{n+1,s})
\end{equation}

As before, $u_n$ is a real scalar bosonic field which denotes the displacement of the $n^{th}$ site along the symmetry axis, $p_n$ is the corresponding momentum and $m$ is mass of the CH group, $t_{n+1,n}$ is the hopping amplitude describing the hopping of electron from site $n$ to $n+1$ and $c_{n,s}^\dagger(c_{n,s})$ is the fermionic creation (annihilation) operator that create (destroy) electrons of spin $s$ at site $n$. 
In the SSH model, the lattice potential is given by 
\begin{equation}
V(u_n,u_{n+1}) = \frac{1}{2}K(u_{n+1}-u_n)^2\ ,
\end{equation}
where $K$ is the spring constant of the bonds and $t_{n+1,n}$ is expanded upto first order:
\begin{equation}
t_{n+1,n} = t_0-\alpha(u_{n+1}-u_n)
\end{equation}

With $\alpha=1$, $t_0=\frac{1}{2a}$ and ignoring the spin variable $s$, the hopping term of the SSH Hamiltonian coincides with the BR Hamiltonian.
The existence of two degenerate ground states of the B-R Hamiltonian  was shown for a heterostructure system with 2n+1 atomic planes\cite{selvan2016charge}, n being the number of pairs of oppositely charged planes with charge fractionalization at the end points \cite{jackiw1981solitons}.

The first two terms of Eq.(\ref{ham}) can be suitably taken as the space and time derivative in the continuum field approximation.
As the potential has a $Z_2$ symmetry, the $\lambda \phi^4$ theory captures the characteristics of the phonon field, with the Lagrangian density,
\begin{equation}\label{lagrangian}
    \mathfrak{L} = \frac{1}{2}\dot{\phi}^2 - \frac{\hbar^2}{2m}\phi'^2+\frac{1}{2}\mu \phi^2 - \frac{\lambda}{4}\phi^4\ ,
\end{equation}
where $\dot{\phi}$ and $\phi'$ denote time and space derivatives respectively.
The energy functional for the phonon field is given by
\begin{equation}\label{energy}
E = \int dx \Bigg(\frac{\hbar^2}{2m}\phi'^2+\frac{\lambda}{4}\phi^4-\frac{\mu}{2} \phi^2 \Bigg)\ 
\end{equation}
with the static field equation
\begin{equation}\label{eom1}
    \frac{\hbar^2}{2m}\phi'' +\mu\phi-\lambda\phi^3=0
\end{equation}
The non-trivial constant solutions, 
\begin{equation}\label{const-sol}
\phi_0 = \pm\sqrt{\frac{\mu}{\lambda}}
\end{equation}
describing the two degenerate ground states of the spontaneously broken $Z_2$ symmetry
has energy
\begin{equation}\label{zeroenergy}
E_0 = -\frac{\mu^2}{4\lambda}L
\end{equation} 
where $L$ is the system size.
The equation permits more general real solutions as well.
Rewriting Eq.(\ref{eom1}) as
\begin{equation}\label{eom2}
\phi''+c_1\phi-c_2\phi^3=0
\end{equation} 
with $c_1 = \frac{2m\mu}{\hbar^2}$ and $c_2=\frac{2m\lambda}{\hbar^2}$, the general solutions are the Jacobi elliptic sine function, given as 
\begin{equation}\label{gen}
    \phi(x) = A(k) sn\Big(\frac{x-x_0}{\alpha(k)};k\Big)
\end{equation}
where $x_0$ is a real constant that can be fixed to 0 without any loss of generality, k is the elliptic modulus $(0\leq k \leq 1)$ and the real parameters A and $\alpha$ are given by 
\begin{equation}\label{ak}
    A^2(k) = \Big(\frac{2k^2}{1+k^2}\Big)\frac{c_1}{c_2}\ \text{and}\ \alpha^2(k) = \frac{1+k^2}{c_1}
\end{equation}
The $sn$ function is a periodic function with period $4K(k)$, where $K(k)$ is the complete elliptic integral of the first kind \cite{abramowitz1964handbook}.
The parameter $A(k)$ is the amplitude of the function and $\alpha(k)$ becomes a suitable length scale whose value will depend on the system under study.
We have one more parameter, $k$ to be linked to some microscopic degree of freedom in the system.
In our case we connect the elliptic modulus to the anharmonic shift parameter $\delta$.

In the continuum limit, all the lattice points are brought sufficiently close to each other leading to an approximate result,
\begin{equation}\label{cont-del}
    \phi(x+2\delta) - \phi(x) \approx \frac{1}{2a}
\end{equation}
which gives 
\begin{equation}\label{taylor}
    \delta \approx \frac{1}{4a\phi'(0)}
\end{equation}
on expansion upto the first order.
Explicitly,
\begin{equation}\label{delta}
    \delta = \frac{k}{1+k^2}\frac{\hbar }{2a\mu}\sqrt{\frac{\lambda}{m}}
\end{equation}

We impose the periodic boundary condition of $\phi(\frac{L}{2})=\phi(-\frac{L}{2})$ to keep the system topologically charge neutral.
It is to be noted that for the consideration of phonon energy, we have excluded  the neutral modes of the end points. 
Therefore this condition on the phonon field is well justified.

Defining the period of oscillation as $4nK(k) = \frac{L}{\alpha(k)}$, where $n$ is the number of pairs of alternating charged atomic planes in the heterostructure, the oscillatory function, sn(x;k) captures the essential features of the alternate distortion of the charged planes of opposite polarity in the system.
Let the state containing $n$ pairs of alternating charged planes be $\phi_{2n}$.
The energy stored in this state can be calculated from Eq.(\ref{energy}).
\begin{equation}\label{2n}
E_{2n} = \int \limits_{-\frac{L}{2}}^{\frac{L}{2}} dx (\frac{\hbar^2}{2m}\phi_{2n}'^2+\frac{\lambda}{4}\phi_{2n}^4 -\frac{\mu}{2}\phi_{2n}^2\Bigg)
\end{equation}
Substituting Eq.(\ref{gen}) into Eq.(\ref{2n}), we get:
\begin{equation}
E_{2n} = -\frac{\hbar^2 A^2(k) k^2}{2m\alpha^2(k)}I(k)
\end{equation}
where 
\begin{equation}
I(k) = \frac{(2+k)}{3k^2}\frac{L}{\alpha(k)} - \frac{2n\pi(1+k)}{3k^2}
\end{equation}

\noindent
Taking the thermodynamic limit, i.e., $L\rightarrow \infty$ or equivalently $k\rightarrow 1$, the energy density is found to be 
\begin{equation}\label{thermo}
\mathfrak{E}_{2n}=\frac{E_{2n}}{L} = -\frac{\mu^2}{4\lambda}
\end{equation}
It is observed that the energy density is same as in the case of the constant solution.
This indicates that the non-uniform state is more likely to form induced by lattice variations.

\vspace{5mm}
In conclusion, we have shown the common origin of the fermion number fractionalization on $LaAlO_3/SrTiO_3$ heterostructure and polyacetylene.
The periodic Jacobi elliptic sine wave solution of the anharmonic lattice description effectively captures the alternate distortion of the charged planes of opposite polarity in $LaAlO_3/SrTiO_3$ heterostructure.
The opposite end points carry fractional charges, as observed in the experiments, well described by the Bell - Rajaraman model.
The charged soliton pairs emerge as a limiting case, where the lattice distortion are localized near the soliton sites, leaving the rest of the structure in a uniform state described by the $\lambda \phi^4$ double well potential.

\bibliographystyle{unsrt}
\bibliography{References}

\end{document}